\documentclass[12pt]{iopart}
 \expandafter\let\csname equation*\endcsname\relax
 \expandafter\let\csname endequation*\endcsname\relax

\usepackage{amssymb, amsmath}

\DeclareMathOperator{\diag}{diag}
\DeclareMathOperator{\sgn}{sgn}

\DeclareMathOperator{\im}{Im}

\newtheorem{thm}{Theorem}
\newtheorem{prop}[thm]{Proposition}

\begin{document}

\title{Universal $K-$ matrix distribution  in $\beta=2$ Ensembles of Random Matrices}

\vskip 0.2cm
\author{Y.~V.~Fyodorov, B.~A.~Khoruzhenko,  and A. Nock}
\address{Queen Mary University of London, School of Mathematical Sciences, London~E1~4NS, United Kingdom}

\begin{abstract}
The $K-$matrix, also known as the ``Wigner reaction matrix'' in nuclear scattering or ``impedance matrix'' in the electromagnetic wave scattering, is given essentially by an $M\times M$ diagonal block of the resolvent $(E-H)^{-1}$ of a Hamiltonian $H$. For chaotic quantum systems the Hamiltonian $H$ can be  modelled by random Hermitian $N\times N$ matrices  taken from invariant ensembles with the Dyson symmetry index $\beta=1,2$ or $4$. For $\beta=2$ we prove by explicit calculation a universality conjecture by P. Brouwer ({\it Phys. Rev. B} \textbf{51} (1995), 16878-84) which is equivalent to the claim that  the probability distribution of $K$, for a broad class of invariant ensembles of random Hermitian matrices $H$, converges to a matrix Cauchy distribution with density ${\cal P}(K)\propto \left[\det{({\lambda}^2+(K-{\epsilon})^2)}\right]^{-M}$  in the limit $N\to \infty$,  provided the parameter $M$ is fixed and the spectral parameter $E$ is taken within the support of the eigenvalue distribution of $H$. In particular, we show that for a broad class of unitary invariant ensembles of random matrices finite diagonal blocks of the resolvent are Cauchy distributed.  The cases $\beta=1$ and $\beta=4$ remain outstanding.
\end{abstract}


\maketitle
The phenomenon of chaotic resonance scattering of quantum waves (or their classical analogues) has attracted considerable theoretical and experimental interest for more than two decades, see e.g. articles in \cite{chao05}. The resonances manifest themselves via fluctuating structures in scattering or transport observables, and statistical properties of such objects can be successfully described within the framework of the Random Matrix Theory \cite{Mel04,FSav11}. The most important object in such an approach is the energy-dependent $M\times M$ random unitary scattering matrix $S(E)$, $S^{\dagger}(E)S(E)=1_M$ which relates amplitudes of incoming and outgoing waves. Here the integer $M$ stands for the number of open channels at given energy, the dagger denotes the Hermitian conjugation and $1_M$ is the $M\times M$ identity matrix. Statistical properties of scattering observables considered at a fixed energy $E$ of incoming waves can be inferred from the corresponding probability density of $S=S(E)$ derived starting from rather general physical principles. Those include unitarity, causality and (if relevant) the time-reversal invariance imposed on $S$  combined with the assumption of maximal entropy (minimum information). The procedure yields
 the so-called {\it Poisson's kernel distribution} wih density \cite{Mel85}:
\begin{equation}\label{poisson}
  P_{\overline{S}}(S)  =\frac{1}{C_\beta}\left| \frac{\det[1_M- \overline{S}^{\dag}\overline{S}]}{
\det[1_M- \overline{S}^{\dag}S]^2}\right|^{(\beta M +2-\beta)/2},
\end{equation}
where $\overline{S}$ stands for the mean of the scattering matrix, $\beta=1,2,4$ is the parameter related to underlying symmetries with respect  to time reversal and $C_\beta$ is a normalization constant. The mean $\overline{S}$ is determined by the details of coupling of the systems to continuum
and thus contains all information which should be specified for a given scattering system. In particular, for the simplest, yet most fundamentally important ``perfect coupling'' case $\overline{S}=0$, and the density in (\ref{poisson}) is constant, implying that the $S-$matrix is uniformly distributed over the unitary matrices of given symmetry. 

Although the above method has proved to be very successful in the statistical description of scattering characteristics at fixed energy \cite{Mel04}, it can not be used to study statistics of fluctuations of the scattering observables over an energy interval comparable with a typical separation between resonances. The latter task can be most successfully achieved in an alternative powerful approach going back to the pioneering work \cite{Ver85} which is based on the paradigm of random matrix properties of the underlying Hamiltonian $H$ describing quantum chaotic behaviour of the closed counterpart of the scattering system.
In such an approach the resonance part of the $S$-matrix is expressed in terms of the resolvent of such a Hamiltonian as
\begin{equation}\label{S1}
  S(E) = \frac{1_M-iK(E)}{1_M+iK(E)} \, , \qquad K(E) =   \textstyle W^{\dag} (E-H)^{-1} W \,,
\end{equation}
where $W$ is an $N{\times}M$ matrix of energy-independent coupling amplitudes between $N$ energy levels of the closed system and $M$ open scattering channels. To study quantum chaos-induced fluctuations of $S$ one then replaces the Hamiltonian $H$ with a random matrix taken from one of the standard random matrix ensembles, usually Gaussian Unitary (GUE, $\beta=2$) if one is interested in the systems with broken time reversal invariance or Gaussian orthogonal (GOE, $\beta=1$) if such invariance is preserved, the case $\beta=4$ being relevant for systems with spin-orbit scattering. The approach proved to be extremely successful, and quite a few scattering characteristics were thoroughly investigated in that framework in the last two decades, mainly by the supersymmetry method \cite{Ver85}-\cite{Fyo97a}. The results of such calculations are found in general to be in good agreement with available experiments
in chaotic electromagnetic resonators (``microwave billiards'') or acoustic reverberation cameras, see e.g. \cite{Kuh05,Hem05,Die10}
and most recently in \cite{Kum13}, as well as with numerical simulations of scattering in such paradigmatic model as quantum chaotic graphs \cite{Kot00}.

The two random matrix approaches described above look very different in their formulation, yet they are meant to describe precisely the same object,
the $S-$matrix for a chaotic system. The consistency therefore requires that the Poisson kernel distribution (\ref{poisson}) for $S$  must follow from the law of distribution of $H$ entering the relation (\ref{S1}). Surprisingly, a direct verification of such a correspondence turns out to be a rather challenging task. The challenge here is that the two objects are related via the resolvent-like $K-$matrix, and to convert the law of distribution of $H$ into that of the resolvent is not at all trivial. A very elegant indirect way round this problem was discovered by P. Brouwer \cite{Bro95} who proposed to choose $H$ from the Cauchy ensemble of random matrices with density 
$P(H)\propto \det{\left[\lambda^2+(H-\epsilon)^2\right]}^{-(\beta N+2-\beta)/2} $, where $\lambda,\epsilon$ are two real parameters.
The main advantage of such a choice is that the resolvent $(E-H)^{-1}$  is Cauchy-distributed  as well albeit with modified parameters and, moreover, diagonal blocks of Cauchy matrices have again closely related distributions. Using these facts Brouwer indeed was able to demonstrate the validity of the Poisson kernel for such a choice of $H$ for all values of $\beta=1,2,4$. He then showed that in the large-$N$ limit the eigenvalue correlation functions in the Cauchy ensemble (called ``Lorentzian'' ensemble by Brouwer) have the standard Dyson form, and conjectured that such equivalence of eigenvalue correlation functions should be enough to ensure the same $S-$matrix distribution is to be shared by all representatives of the corresponding universality class. Although such conjecture sounds very natural, the particular mechanism by which the generic spectral properties of $H$ are translated into universality of the probability density of the $K-$matrix and then $P(S)$ remained unclear. To the best of our knowledge no further attempts to verify universality of the $S-$matrix distribution were undertaken in the literature apart from (i) the simplest case $M=1$ and $H\in GUE$ considered in \cite{Fyo97a} and (ii) the recent work \cite{Man10} which however concentrated on the universality of two-point spectral correlations of the individual $S-$matrix entries rather than on the one-point matrix distribution.

In the present work we establish the universality of the Poisson distribution for the $S$-matrix under the condition of equivalent coupling to continuum in all scattering channels. Since the $K$- and $S$-matrices are related via the Cayley transformation (\ref{S1}), the claim of the Poisson kernel distribution for the $S-$matrix is equivalent to claiming that the $K$-matrix is Cauchy with density 
\begin{equation}\label{cauchy}
   P_{\beta}(K)\propto \left[\det{({\lambda}^2+(K-{\epsilon})^2)}\right]^{-\frac{1}{2}(\beta M+2-\beta)},
\end{equation}
where the width ${\lambda}$ and the mean ${\epsilon}$ of the Cauchy distribution  are determined by the mean $\bar S$ of the $S$-matrix and vice versa. Such equivalence can be verified by transforming to the eigenvalues of the $S$-and $K$-matrices. 

In the present work, under fairly generic assumptions on $H$, we verify that the law of distribution of the $K$-matrix is indeed Cauchy and relate its parameters $\lambda$ and  $\epsilon$ to the strength of the coupling amplitudes $W$ and the density of states of the underlying matrix $H$ as well as details of its (invariant) distribution. We would like to note that the statistical characteristics of the $K$-matrix are directly accessible in microwave experiments \cite{Hem05} where it is related to the real part of the {\it impedance} in the regime of small losses. Our paper stems from attempts  to understand better the mechanism behind universality of the probability distribution of  finite blocks of random matrix resolvents and to provide an {\it ab initio} explicit derivation of this distribution for generic invariant ensembles of random matrices. 


We will start with showing that the universality of the $K$-matrix in question follows from the universal limit of a very general spectral object -- the product of the ratios of powers of characteristic polynomials $\det(E-H)$ of random matrices $H$.

To that end it is necessary to mention that  in the literature there exist two alternative choices of the matrix  $W$ 
of coupling amplitudes in (\ref{S1}). The standard choice is to follow the original paper \cite{Ver85} and to consider the columns ${\bf w}_1,\ldots {\bf w}_M$ of $W$ as mutually orthogonal $N-$component vectors, real for $\beta=1$ and complex for $\beta=2$.
The case of equivalent channels then corresponds to 
$({\bf w}_a,{\bf w}_b)=\gamma \delta_{ab}$ for all $a,b=1,\ldots M$, or, equivalently, 
\begin{equation}\label{FA}
   W^{\dagger}W= \gamma 1_M,
\end{equation}
where $\gamma>0$ is a coupling constant. An alternative choice which was suggested originally in \cite{SokZel89} is to consider the columns of $W$ to be independent Gaussian vectors with joint probability density function 
\begin{equation}\label{RA}
P( W)\propto e^{-\frac{\beta N}{2 \gamma}\Tr  W^{\dag}W}, \quad \langle W^{\dagger}W\rangle = \gamma 1_M.
\end{equation}
Both choices are expected to lead to the same results in the limit $ N\to \infty$ as long as the number of channels $M$ remains fixed. Such an equivalence was explicitly verified in \cite{Lem95} for particular scattering characteristics (Wigner delay times), but is expected to hold generally. 

We shall first consider  the random amplitude case (\ref{RA}) and show the equivalence to the fixed amplitude case (\ref{FA}) for $\beta=2$ at the end of the paper. For notational convenience, it is more convenient to work with the rescaled $K$-matrix
\begin{equation*}
  \tilde{K}=\frac{K}{\pi \rho(E)}, \quad K=W^{\dag}(E-H)^{-1}W,
\end{equation*}
rather than the $K$-matrix itself, with $\rho(E)$ being the large-$N$ limit of the mean eigenvalue density of $H$ at point $E$ inside the support of $\rho(E)$ (so that $\rho(E)>0$).

We start with the representation 
$P(\tilde{K})=\int{\cal F}_{\beta, N}(X)\exp({i\frac{\beta}{2}\Tr 
 X\tilde{K}})\, dX $ 
of the probability density function of $\tilde K$ in terms of the characteristic  function 
\begin{equation}\label{F}
  {\cal F}_{\beta, N}(X)=\left\langle\exp{\left(-i\frac{\beta}{2}\Tr X\tilde{K}\right)}\right\rangle, 
\end{equation}
where the matrix $X$ has the same dimensions and symmetries as the matrix $\tilde K$ and  the angular brackets stand for the averaging over all random variables the matrix $\tilde{K}$ depends on. Writing $\tilde{K}= \frac{1}{\pi \rho(E)} W^{\dag}RW, \, R=(E-H)^{-1}$, we first  perform the averaging over the coupling matrix $W$ in (\ref{F}) which amounts to performing a Gaussian integral over the vectors ${\bf w}_c$:
\begin{equation}\label{1}
 \int P(W) \exp\left(-\frac{i\beta}{2 \pi \rho(E)}\Tr XW^{\dag}RW\right) dW
\,  = \,  \prod_{c=1}^M\left[\det\left(1_N+\frac{i \gamma x_c}{\pi N \rho(E)}R\right)\right]^{-\frac{\beta}{2}}\, .
\end{equation}
Here $x_1,\ldots, x_M$ are the eigenvalues of the Hermitian $M\times M$ matrix $X$
and $dW$ stands for the appropriately normalized Lebesgue measure on the space of complex or real $N\times M$ matrices $W$. The easiest way to verify (\ref{1}) is by diagonalizing $X=T\,\diag (x_1,\ldots,x_M)\,T^{-1}$ where $T$ is orthogonal for $\beta=1$ and unitary for $\beta=2$.
Then one changes $(TW)\to W$ and exploits the invariance of $W^{\dag}W$ and the measure $dW$ with respect of such a transformation.
At the next step we bring the  characteristic  function ${\cal F}_{\beta, N}(X)$ to the following form:
\begin{equation}\label{2}
  {\cal F}_{\beta, N}(X)=\left\langle\prod_{c=1}^M \frac{\left[\det{\left(E-H\right)}\right]^{\frac{\beta}{2}}}
{[\det{(E+\frac{i \gamma x_c}{\pi N \rho(E)}-H)}]^{\frac{\beta}{2}}}\right\rangle_{\!\!\!H}\, ,
\end{equation}
where the angular brackets now stand for the averaging over the $N\times N$  matrices $H$.
The above relation is {\it exact} in the random amplitude model (\ref{RA})  for any choice of $N$ and $M$. We will show at the end of the paper that for $\beta=2$ the same equation (\ref{2}) is valid asymptotically in the fixed amplitude model (\ref{FA}) in the limit $N\gg M$ provided the probability density of $H$ is rotationally invariant.

For $\beta=2$ and $\beta=4$ the object in the right-hand side of (\ref{2}) is well-studied in the Random Matrix Theory \cite{FS03a,FS03b,BS06}.
 In particular, in the simplest case $\beta=2$ the formula (2.14) from \cite{FS03b} appears to be most useful
 for our goals. Namely, for $N\times N$ matrices $H$ distributed according to an invariant ensemble density with polynomial potential $V$,
\begin{equation}\label{H}
  P(H)\propto \exp\left[ {-N \Tr  V(H)} \right], \quad V(H)=\sum_{l=0}^p c_l H^{2l}, \quad c_p>0, 
\end{equation}
the following universal relation holds asymptotically\footnote{We restrict ourselves to the polynomial potentials in (\ref{H}) for the notational convenience. The asymptotic relation (\ref{9}), and as a consequence our Proposition \ref{p1} hold for invariant ensembles of random matrices under fairly general conditions on the matrix measure, see the recent paper \cite{BS12} }:
\begin{equation}
\begin{aligned} \label{9}
 & \lim_{N\to \infty}\left\langle \prod_{c=1}^M \frac{\det(E+\eta_c/(N\rho(E))-H)}{\det(E+\zeta_c/(N\rho(E))-H)} \right \rangle_H \\
& = (-)^{M(M-1)/2} \exp \left(-\pi \alpha_E \sum_{c=1}^M (\zeta_c- \eta_c) \right) \frac{\Delta\{\zeta,\eta\}}{\Delta^2\{\zeta\}\Delta^2\{\eta\}} \det(S(\zeta_i-\eta_j)),
\end{aligned}
\end{equation}
where $\Delta\{\eta\}=\prod_{i<j}(\eta_i-\eta_j)$ is the Vandermonde determinant, and
\begin{equation}
   S(\zeta-\eta)=\frac{\exp(i \pi \sgn(\im\zeta)(\zeta-\eta))}{\zeta-\eta}, \quad \alpha_E=\frac{V'(E)}{2 \pi \rho(E)}.
\end{equation}
 An analogous result for averaged products of ratios of characteristic polynomials with $\beta=4$ is also known \cite{BS06}, but has a more complex structure, with Pfaffians replacing determinants. Unfortunately,
for $\beta=1$ no result of comparable generality seems to be known for the products of {\it square roots} of the characteristic polynomials , though for $M=1,2$ it can be in fact evaluated in closed form, see e.g. \cite{FyoSav12} and references therein.  Below we consider in full generality only  the case of Hermitian ensembles with $\beta=2$, whereas the cases $\beta=4$ and especially $\beta=1$ remain a challenge to us and are currently under investigation.

With the asymptotic relation (\ref{9}) in hand, one can evaluate the characteristic function (\ref{2}) of the rescaled $K$-matrix in the limit $N\to\infty$ and $M$ fixed.

\begin{prop} \label{p1} Assume that the $N\times N$ matrix $H$ has invariant  distribution (\ref{H}). Then in the random amplitude model (\ref{RA}) we have $\lim_{N\to\infty}  {\cal F}_{\beta=2, N}(X) = {\cal F}_{\beta=2}(X)$, where 
\begin{equation} \label{14}
 {\cal F}_{\beta=2}(X) = (-)^{M(M-1)/2}\, \frac{\exp[-i\gamma \alpha_E\Tr X]}{\Delta\{X\}}  
\det \begin{pmatrix}
 g_{M-1}(x_1) & \ldots  & g_{M-1}(x_M)\\
\vdots & \ddots & \vdots \\
   g_0(x_1)  & \ldots & g_0(x_M) \end{pmatrix},
\end{equation}
with
\begin{equation} \label{14a}
  g_{M-n}(x)=\exp\left(-\gamma |x| \right) x^{n-1} \sum_{l=0}^{M-n} \frac{1}{l!} |\gamma x|^{l}.
\end{equation}
\end{prop}
\textbf{Proof.}  
To adjust equation \eqref{9} to our goals we first set $\eta_1=\eta_2=\ldots=\eta_M=0$ there. In this limit  $\Delta\{\zeta,\eta\}/(\Delta\{\zeta\}\Delta\{\eta\}) \rightarrow (\zeta_1 \times \ldots \times \zeta_M)^M$ and equation \eqref{9} becomes
\begin{equation}
 \begin{aligned} \label{limit1}
 &\lim_{N\to \infty}\left\langle \prod_{c=1}^M \frac{\det(E-H)}{\det(E+\frac{\zeta_c}{N\rho(E)}-H)} \right \rangle_H
 = \exp\left(\sum_{c=1}^M \pi \zeta_c \big(i\sgn(\im\zeta_c)-\alpha_E \big) \right) \\
 &\times (-)^{M(M-1)/2} \frac{(\zeta_1 \times \ldots \times \zeta_M)^M}{\Delta\{\zeta\}}
  \lim_{\eta_1 \dots \eta_M \to 0} \frac{1}{\Delta\{\eta\}} \det \begin{pmatrix}
 \tilde{g}(\zeta_1,\eta_1) & \ldots &  \tilde{g}(\zeta_1,\eta_M)\\
\vdots & \ddots & \vdots \\
   \tilde{g}(\zeta_M,\eta_1)  & \ldots & \tilde{g}(\zeta_M,\eta_M) \end{pmatrix}
 \end{aligned}
\end{equation}
where
\begin{equation}
  \tilde{g}(\zeta,\eta)=\frac{\exp[-i\pi \sgn(\im \zeta) \eta]}{\zeta-\eta}.
\end{equation}
The limits are now performed successively applying the L'Hospital's rule, the final result for the second line in equation \eqref{limit1} being
\begin{equation}
\lim_{\eta_1 \dots \eta_M \to 0} \frac{1}{\Delta\{\eta\}} \det \left[\tilde{g}(\zeta_i,\eta_j) \right]_{1 \leq i,j \leq M}
=
\left( \prod_{n=1}^{M-1} \frac{1}{n!} \right) \det \begin{pmatrix}
 \tilde{g}_{M-1}(\zeta_1) & \ldots  & \tilde{g}_0(\zeta_1)\\
\vdots & \ddots & \vdots \\
   \tilde{g}_{M-1}(\zeta_M)  & \ldots & \tilde{g}_0(\zeta_M) \end{pmatrix},
\end{equation}
where we have defined
\begin{equation}
   \tilde{g}_n(\zeta)= \left. \frac{\partial^n}{\partial \eta^n}  \tilde{g}(\zeta,\eta) \right|_{\eta=0}
=\sum_{l=0}^n \frac{n!}{(n-l)!}[-i \pi \sgn(\im \zeta)]^{n-l} \zeta^{-l-1}.
\end{equation}
Finally,  by redefining $g_n(\zeta)= e^{i\pi\zeta\sgn (\im \zeta)} \zeta^M \tilde{g}_n(\zeta)/n!$, several factors in front of the determinant can be absorbed into the determinant. After identifying $\zeta_c \to i \gamma x_c/\pi$ we arrive at (\ref{14}). This completes our proof of Proposition \ref{p1}. \hfill $\Box$
\vspace{0.5\baselineskip}

At the next step we observe that achieving our main goal is equivalent to verifying that  ${\cal F}_{\beta=2}(X)$ is the characteristic function of a matrix Cauchy distribution.

\begin{prop} \label{p2}
\begin{equation}\label{4}
  \int
\frac{e^{-i\Tr {K}X}\, d {K}}{\det{[\gamma^2+({K}-\gamma \alpha_E)^2]^M}} = \frac{\pi^M M!}{\gamma^{M^2} 2^{M(M-1)}} {\cal F}_{\beta=2}(X)\,,
\end{equation}
where the integral is over the set of all Hermitian $M\times M$ matrices ${K}$. 
\end{prop}
\textbf{Proof.} A standard random matrix calculation which involves changing the variables of integration in \eqref{4}  to the eigenvalues and eigenvectors of $K$ and then applying the  Itzykson-Zuber-Harish-Chandra (IZHC) formula, see e.g.  \cite{Mor11}, and the Andr\'eief-de Bruijn integration formula yields 
\begin{equation}\label{5}
 \int\frac{e^{-i\Tr {K}X}\ d{K}}{\det[\gamma^2+({K}-\gamma \alpha_E)^2]^M} = \left(\prod_{n=1}^M n! \right)
 \frac{e^{-i\gamma \alpha_E \Tr X}}{\Delta\{X\}}
 \det \begin{pmatrix}
        f(x_1)  & \ldots  & f(x_M)\\
        f'(x_1) & \ldots & f'(x_M) \\
        \vdots  & \ddots & \vdots \\
        f^{(M-1)}(x_1) & \ldots & f^{(M-1)}(x_M)
      \end{pmatrix}\, .
\end{equation}
Here 
\begin{equation}\label{6}
f(x)=\int_{-\infty}^{\infty}d{k}\  \frac{e^{-i{k}x}}{(\gamma^2+{k}^2)^M} = \sqrt{\tfrac{2}{\pi}} c_{\gamma} |\gamma x|^{M-\frac{1}{2}}K_{M-\frac{1}{2}}(|\gamma x|), \quad f^{(M)}(x)=\frac{d^M}{dx^M}f(x),
\end{equation}
where $K_{\nu}(x)$ is the modified Bessel (Macdonald) function and the constant is given by $c_{\gamma}=\frac{\pi}{\gamma^{2M-1}2^{M-1}\Gamma(M)}$. In particular, for $M=1$ we have $f(x)=c_{\gamma} e^{-\gamma|x|}$, for higher $M$ we have
\begin{equation}\label{7}
f_M(x) =c_{\gamma} e^{-\gamma|x|} \sum_{l=0}^{M-1} \frac{(M-1+l)!}{l!(M-1-l)!2^l}|\gamma x|^{M-1-l},
\end{equation}
where we added a subscript to indicate the $M$-dependence. Using a recursive relation for the derivative of the Macdonald function we can show that  $f_{M}'(x)= -\gamma x f_{M-1}(x)$. Inductively one gets for higher derivatives
\begin{equation} \label{8}
 f_{M}^{(m)}(x)=\sum_{l=0}^{\lfloor m/2 \rfloor} \frac{m! (-1)^{m-l}}{l!(m-2l)!2^l} (\gamma x)^{m-2l}f_{M-m+l}(x),
\end{equation}
where $\lfloor \cdot \rfloor$ denotes the floor-function. This enables us to simplify the determinantal structure of \eqref{5}.  By successively adding to the $n$-th row appropriate linear combinations of all preceding rows $1,2,\ldots, n-1$, and exploiting yet another recursive relation for the Macdonald function one can remove all terms in the equation \eqref{8} but the one for $l=0$, leading to
\begin{equation} \label{16}
 \det \begin{pmatrix}
        f_{M}(x_1)  & \ldots  & f_{M}(x_M)\\
        f_{M}'(x_1) & \ldots & f_{M}'(x_M) \\
        \vdots  & \ddots & \vdots \\
        f_{M}^{(M-1)}(x_1) & \ldots & f_{M}^{(M-1)}(x_M)
      \end{pmatrix}
\propto
 \det \begin{pmatrix}
        f_{M}(x_1) & \ldots  & f_{M}(x_M)\\
        x_1 f_{M-1}(x_1) & \ldots & x_M f_{M-1}(x_M) \\
        \vdots & \ddots & \vdots \\
        x_1^{M-1} f_1(x_1)  & \ldots & x_M^{M-1} f_1(x_M)
      \end{pmatrix},
\end{equation}
where the proportionality constant is $(-)^{M(M-1)/2} \left(\prod_{m=1}^M \frac{(M-1)! \gamma^{M-1} 2^{m-1}}{(2m-2)!} \right)$ and the combination involved in the $n$-th row in the right-hand side is given explicitly by
\begin{equation} \label{17}
 x^{n-1} f_{M-n+1}(x) =c_{\gamma}  e^{-\gamma|x|} x^{n-1} \sum_{l=0}^{M-n} \frac{(2M-2n-l)!}{l!(M-n-l)!2^{M-n-l}}|\gamma x|^l.
\end{equation}
The equations \eqref{14a} and \eqref{17} have a very similar structure, though the coefficients of the terms in the sum are still different. In fact this similarity can be further exploited to show that the determinants in equations \eqref{14} and \eqref{5} (or equivalently \eqref{16}) are proportional to each other, thus verifying the equation \eqref{4}.

 We start our demonstration of this fact with bringing the first row of the determinant in equation \eqref{14} to the form coinciding with the first row of the determinant in \eqref{16}. Since the zeroth and the first order coefficients of $g_{M-1}(x)$ are both equal to unity, and the two corresponding coefficients are also equal in the expression for $f_M(x)$ (but are different from unity) we can safely change those coefficients in $g_{M-1}(x)$ to the coefficients in $f_M(x)$ as such a change  gives rise to a constant proportionality factor for the determinant.

 The main observation is that the adjustment of both the coefficients $a_n$ and $a_{n+1}$, given that all previous coefficients are already adjusted, can be done simultaneously by adding the $(2n+1)-$th row multiplied with the factor
\begin{equation} \label{18}
 c_n=(-1)^n\frac{(2M-2n-2)!}{n!(M-n-1)!2^{M-1}}.
\end{equation}
 For this procedure to work we need to verify, for any integer $n$, the following identity: 
\begin{equation} \label{19}
 \sum_{l=0}^n(-1)^l\frac{(2M-2l-2)!}{l!(M-l-1)!2^{M-1}} \frac{1}{(2n+\delta-2l)!} = \frac{(2M-2n-\delta-2)!}{(2n+\delta)!(M-2n-1)!2^{M-2n-\delta-1}},
\end{equation}
with $\delta=0$ or $\delta=1$. The left-hand side of (\ref{19}) is what becomes of the ($n+\delta$)-th order coefficient of $g_{M-1}(x)$ after adding multiples of all odd rows up to $2n+1$, choosing the multiplication factors according to \eqref{18}.
The right-hand side  equals to the corresponding ($n+\delta$)-th order coefficient of $f_M(x)$. 
Both equations can be conveniently combined into a single relation:
\begin{equation}\label{rela}
 \sum_{l=0}^{\lfloor m/2 \rfloor} (-1)^l \binom{2\mathcal{M}-2l}{m-2l} \binom{\mathcal{M}}{l} = 2^m \binom{\mathcal{M}}{m},
\end{equation}
where $\mathcal{M}=M-1$ and $m=2n$ or $m=2n+1$. To verify (\ref{rela}) we first express the first binomial on the left-hand side by a contour integral using its generating function and the Cauchy's residue theorem. The summation over $l$ is then performed in the integrand using the binomial theorem, and the resulting contour integral can be again evaluated by the residues, yielding precisely the right-hand side of the relation (\ref{rela}).

 We conclude that it is indeed possible to transform $g_{M-1}(x)$ into $f_M(x)$ by adding multiples of all odd rows to the first row. Note that in each step two coefficients get adjusted simultaneously, and this is precisely the mechanism ensuring the whole procedure being functional. Had it not been for that property, we would be only able to change half of the coefficients to the required form, since adding even rows to odd rows or vice versa is meaningless due to their rather different structure. 
 All remaining odd rows as well as all even rows can be treated by exactly the same procedure, since the coefficients involved
are essentially the same as before.
Note also that as the very last row contains on both sides the function $ e^{-\gamma|x|} x^{n-1}$  the coincidence is ensured automatically. This completes our proof of Proposition \ref{p2} except for the proportionality constant. It can be found by considering the $X=0$ case. In that case we have ${\cal F}_{\beta=2}(0)=1$ and the integral on the left-hand side yields the given constant. \hfill $\Box$
\vspace{0.5\baselineskip}

Since the characteristic function uniquely determines the law of distribution, one concludes from Propositions \ref{p1} and \ref{p2} that the distribution of the $K$-matrix (\ref{S1}) converges in the limit $N\to\infty$ to the matrix Cauchy distribution with density $P_{\beta=2}(K)$ (\ref{cauchy})
having mean $\epsilon =\gamma V^{\prime}(E)/2$ and width $ \lambda=\pi\gamma\rho(E)$. This corresponds to the Poisson kernel distribution \eqref{poisson} for the $S$-matrix with mean $\overline{S}_{ij}=\frac{1-\pi \gamma \rho(E)(1+i \alpha_E)}{1+\pi \gamma \rho(E)(1+i \alpha_E)} \delta_{ij}$.  The case of perfect coupling is then obtained for $\alpha(E_{max})=0$ and $\pi \gamma \rho(E_{max})=1$, where $E_{max}$ denotes the point where $\rho(E)$ has its maximum. Thus indeed, the Poisson kernel distribution for the $S$-matrix is universal in the random amplitude model (\ref{RA}) in that it does not depend on the choice of the random matrix ensemble for the underlying matrix $H$. 

Finally, we would like to demonstrate that the fixed amplitude model (\ref{FA}) 
yields the same universal behaviour of the $K-$matrix in the limit $N\to \infty$.
Let us again consider the characteristic function ${\cal F}_{\beta=2,N}(X)=\left\langle\exp[{-i\Tr (\frac{1}{\pi \rho(E)}XW^{\dag}RW )}]\right\rangle_{H}=\left\langle\exp({-i\Tr \Gamma_x UR_{\Lambda}U^{\dagger}})\right\rangle_{H}$, with $\Gamma_x=WXW^{\dagger}$ and $R_{\Lambda}=\left[\pi \rho(E)(E-\Lambda)\right]^{-1}$. Here $U$ is the unitary matrix of eigenvectors of $H$ and $\Lambda=\diag \{\lambda_1,\ldots,\lambda_N\}$ stands for the diagonal matrix of the corresponding eigenvalues. The averaging over $H$ then can be performed in two steps, the first step being the averaging over the Haar measure on the unitary group $U(N)$. As this is  again a special case of the IZHC integral it can be done explicitly. The important new feature however is that the $N\times N$ matrix $\Gamma_x$ is of a reduced rank, with its $M\ll N$ nonzero eigenvalues coinciding with eigenvalues $\gamma x_c,\, c=1,\ldots, M$ of the matrix $XW^{\dagger}W=\gamma X$, the rest of $N-M$ eigenvalues being exactly zero. At the same time the resolvent matrix $R_{\Lambda}$ is of  the full rank $N$. The problem of performing
the IZHC integral for two matrices of different rank can be most efficiently done by employing equation (A4) of the Appendix A in the paper \cite{FS2003}  (which is in fact closely related to the so-called duality IZHC relation, see equation (17.3.8) in \cite{Mor11}). In our case it takes the form:
\begin{equation}\label{IZHCfinrank1}
\left\langle\exp\left({-i\Tr \Gamma_x UR_{\Lambda}U^{\dagger} }\right) \right\rangle_{U}\propto \frac{\det{X}^{M-N}}{\Delta\{X\}} \int_{{\cal C}_{\Gamma}}  \Delta\{Y\} \prod_{c=1}^M \frac{e^{-i \gamma x_c y_c}} {\det{\left(y_c-R_{\Lambda}\right)}}\, dy_1\ldots y_M
\end{equation}
where the integration goes over the complex variables $y_1,\ldots, y_M$ along contours parallel to the real axis such that  $\sgn(\im y_c)=-\sgn (x_c)$. The proportionality constant is given by $\prod_{c=1}^M (-2 \pi i)(-i\gamma)^{N-c}/(N-c)!$. Now we should perform the next step of the ensemble average over the eigenvalues $\Lambda$ of $H$ entering via the resolvent $R_{\Lambda}$. After rescaling $y_c\to Ny_c$ and a simple rearranging in the integrand we can see that the eigenvalue-averaged right-hand side of (\ref{IZHCfinrank1}) is proportional to
\begin{equation}\label{IZHCfinrank2}
 \frac{\det{X}^{M-N}}{\Delta\{X\}} \int_{{\cal C}_{\Gamma}}  \Delta\{Y\} e^{-N\sum_{c}\left(i\gamma x_c y_c+\ln{y_c}\right)} \left\langle \prod_{c=1}^M \frac{\det{(E-\Lambda)}} {\det{\left(E-\frac{1}{\pi N \rho(E) y_c}-\Lambda\right)}}\right\rangle_{\Lambda}\! dy_1\ldots y_M
\end{equation}
In the limit $N\to\infty$ the
integrals over $y_c$ can be straightforwardly evaluated by the saddle-point method, with the saddle-point values
being given by $y^{(s.p)}_c=\frac{i}{\gamma x_c}$. This is justified as equation (\ref{9}) ensures that
the expected value in the integrand tends for $N\to \infty$ to a well-defined limit of the order of unity
along contours in the vicinity of the chosen saddle point. Moreover, one can show that the saddle-point can be reached by deforming the original contours without crossing any singularities of the integrand. Furthermore, $\Delta\{Y^{(s.p)}\}\propto \Delta\{X\} \det{X}^{-(M-1)}$ and the Gaussian fluctuations around the saddle-point value yield the factor $ \det{X}^{N-1}$. Taking all these facts together we see that (\ref{IZHCfinrank2}) indeed reproduces the expression for ${\cal F}_{\beta=2}(X)$ from \eqref{2}, and hence in the fixed amplitude model (\ref{FA}) the $K$-matrix in the limit $N\to\infty$ has the Cauchy distribution  with density (\ref{cauchy}). This result has an interesting corollary. If $\mathbf{w}_c$ are chosen to be the first $M$ columns of the $N\times N$ identity matrix, then $W^{\dag} (E-H)^{-1} W$ is nothing else as the $M\times M$ block of the resolvent $(E-H)^{-1} $. Therefore for invariant ensembles of Hermitian random matrices $H$, finite blocks of the resolvent of $H$ are Cauchy-distributed in the limit of large matrix dimension.

\section*{Acknowledgements}
 A.N. is grateful to Dr. Nick Simm for numerous informative discussions and useful advices.
 Y.V.F. and A.N. were supported by EPSRC grant EP/J002763/1 ``Insights into Disordered Landscapes via Random Matrix Theory and Statistical Mechanics''.
\vspace{\baselineskip}

\end{document}